\documentclass[lettersize,journal]{IEEEtran}
\usepackage{amsmath,amssymb,amsfonts}
\usepackage{cases}

\usepackage{graphicx}
\usepackage[caption=false,font=footnotesize]{subfig} 

\usepackage{array}
\usepackage{booktabs}
\usepackage{makecell}
\usepackage{multirow}

\usepackage{algorithm}
\usepackage{algorithmic}
\usepackage{mathtools}  

\usepackage{stfloats}
\usepackage{xcolor}
\usepackage{textcomp}
\usepackage{url}
\usepackage{verbatim}
\usepackage{cite}
\usepackage{comment}
\usepackage{bm}

\usepackage[colorlinks=true, linkcolor=cyan, citecolor=cyan, urlcolor=cyan]{hyperref}

\pagecolor{lightgray!0}
\color{black}

\begin{document}
\title{Data-Driven Robust Dynamic Wireless Charging Lane Planning 
	with Tunable Profit--Risk Trade-off and Modified User Equilibrium}
\author{Xin Cui,\!~\IEEEmembership{Member,~IEEE},~Yujun Lin,~\IEEEmembership{Member,~IEEE},~Ziang Liu,~\IEEEmembership{Member,~IEEE},\\~Francis Yunhe Hou,~\IEEEmembership{Fellow,~IEEE}, Wei Liu,~\IEEEmembership{Senior Member,~IEEE}%
\thanks{X. Cui, Y. Lin, and Z. Liu are with the Department of Electrical and Computer Engineering, The University of Hong Kong, 999077, Hong Kong, and also with the HKU Shenzhen Institute of Research and Innovation, Shenzhen 518053, China (e-mail: cuixincxcx@163.com; lyj13971224997@163.com; liuziang1119@163.com).}

\thanks{F. Y. Hou is with the Bradley Department of Electrical and Computer Engineering, Virginia Polytechnic Institute and State University, Blacksburg, VA 24061, USA (e-mail: fyhou@vt.edu).}

\thanks{Wei Liu is with the Research Centre for Electric Vehicles and Department of Electrical and Electronic Engineering, The Hong Kong Polytechnic University, Hong Kong (e-mail: wei.liu@polyu.edu.hk) (Corresponding author: Wei Liu).}
\vspace{-0em}
}

\maketitle
\begin{abstract}
Dynamic wireless charging (DWC) has attracted increasing attention as a complement to conventional plug-in charging for electric vehicles (EVs). However, existing planning methods for dynamic wireless charging lanes (DWCLs) overlook realistic EV routing and charging behavior and lack explicit out-of-sample risk control under traffic-demand and renewable-energy uncertainties. To address these limitations, this paper proposes a data-driven two-stage robust planning framework that coordinates DWCL deployment with the operation of coupled traffic--power networks (CTPNs). A data-driven uncertainty set is constructed from joint historical scenarios to preserve spatio-temporal correlations without imposing parametric distributional assumptions. A modified user-equilibrium (MUE) model is proposed to capture realistic EV routing and charging behavior by incorporating state-of-charge (SoC) dynamics and endogenous charging decisions. The resulting two-stage robust model with mixed-integer recourse is reformulated as an equivalent single-stage scenario-approach problem and solved using a customized nested column-and-constraint generation algorithm. An \textit{a posteriori} feasibility guarantee is established to quantify out-of-sample risk. Based on this guarantee, the scenario set is iteratively enriched to meet a prescribed risk tolerance. Case studies on a 69-node power distribution network and a 24-node traffic network demonstrate that the proposed MUE model prevents overinvestment in DWCLs, while the proposed risk-quantification procedure enables a tunable profit--risk trade-off.
\end{abstract}
		

\vspace{-1em}
\section{Introduction}
\begin{figure}[!t]\vspace{-1.3em}
	\centering
	\includegraphics[width=7.2cm]{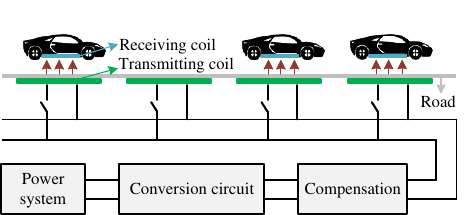}
	\caption{Schematic diagram of dynamic wireless charging.}
	\label{fig1}\vspace{-0em}
\end{figure}

\IEEEPARstart{E}{lectric} vehicles (EVs) provide an environmentally friendly and cost-effective alternative to conventional internal-combustion-engine vehicles \cite{li2022electric}. However, conventional plug-in charging has several limitations, including long charging times, range anxiety, additional land requirements for charging facilities, and manual connection \cite{cui2025modeling}. Dynamic wireless charging (DWC) can address these limitations. Fig.~\ref{fig1} illustrates a DWC system. Through electromagnetic induction between transmitting coils embedded in the roadway and receiving coils mounted on the EV, DWC enables EVs to receive energy while driving \cite{shanmugam2022systematic}. Several DWC projects have been implemented worldwide. For example, Sweden is constructing a permanent dynamic wireless charging lane (DWCL) for EVs \cite{min2023sweden}, while Gumi City in South Korea operates a 15-mile DWCL with an efficiency of approximately 83\% \cite{bansal2015charging}. These projects demonstrate the practical feasibility of DWC and underscore the need for effective DWCL planning.

Existing studies of DWC in coupled traffic--power networks (CTPNs) have primarily focused on post-deployment operation. Wei \emph{et al.} \cite{wei2016optimal} proposed an optimal traffic--power flow model based on static user equilibrium (UE). Building on this, Lv \emph{et al.} \cite{lv2019optimal} extended the static UE model to a semi-dynamic UE formulation that captures traffic evolution across different time intervals for traffic--power coordination. Shi \emph{et al.} \cite{shi2022efficient} designed a framework that integrates DWC systems with real-time electricity markets. Liu \emph{et al.} \cite{liu2024traffic} introduced a tri-level optimization model for toll design in CTPNs with DWCLs. Manshadi \emph{et al.} \cite{manshadi2017wireless} developed a decentralized economic dispatch model for traffic networks (TNs) coupled with power transmission networks. However, these studies assume fixed DWCL locations and do not consider the DWCL investment decisions addressed in this paper.

DWCL planning has been studied more extensively in stand-alone TNs, with attention to road capacity deterioration \cite{he2020optimal}, flow capturing \cite{riemann2015optimal}, electricity pricing \cite{liu2021optimal}, and multi-route traffic operations \cite{hwang2017system}. Because these models omit power-network constraints, their deployment decisions may induce excessive charging demand or voltage violations~\cite{liu2023bilevel,xia2021distributed,zhou2023planning}. DWCL planning models for CTPNs remain limited. Liu \emph{et al.} \cite{liu2023bilevel} proposed a bilevel planning model in which power flow allocation is modeled at the upper level and traffic flow allocation at the lower level to capture EV charging responses. Xia \emph{et al.} \cite{xia2021distributed} formulated a distributed DWCL planning model to maximize system-level economic benefits. Zhou \emph{et al.} \cite{zhou2023planning} developed a planning framework with multi-class UE to co-plan static and dynamic charging facilities in CTPNs.

Despite these advances, existing DWCL planning models do not adequately represent EV traffic assignment, which determines the spatial and temporal distribution of charging loads. System-optimal traffic assignment \cite{liu2023bilevel,xia2021distributed} and user equilibrium (UE) \cite{wei2016optimal,lv2019optimal,shi2022efficient,manshadi2017wireless,liu2024traffic,zhou2023planning} are commonly used for DWCL planning. The former assumes full cooperation among travelers to minimize system travel costs, an assumption that is difficult to realize in practice. The latter represents individual route choice by assuming that each EV selects a minimum-cost route. However, existing UE models have not incorporated both battery state-of-charge (SoC) dynamics and endogenous charging decisions. Models that omit SoC dynamics may produce energy-infeasible routes. Moreover, omitting the endogenous charging decision, i.e., assuming that EVs charge whenever they traverse a DWCL, may lead to an overestimation of charging loads. This is because EVs with sufficient energy may traverse the DWCL without charging. Therefore, it is necessary to develop a modified user-equilibrium (MUE) framework that incorporates SoC dynamics and endogenous charging decisions to estimate route flows and charging loads more consistently.

The second limitation of existing DWCL planning models concerns their treatment of uncertain traffic demand and renewable energy source (RES) output. Stochastic programming (SP), distributionally robust optimization (DRO), and robust optimization (RO) are widely used to address such uncertainties. SP relies on an assumed or empirical probability distribution \cite{ehsan2019active,wu2017stochastic,zeng2019solving}, which may yield poor out-of-sample performance when data are limited or the assumed distribution is misspecified. DRO reduces dependence on a single probability distribution by optimizing over an ambiguity set constructed using divergence measures \cite{xie2018planning,zhou2020planning} or Wasserstein balls \cite{liu2023energy}. However, operational recourse in DWCL planning includes binary decisions, yielding a two-stage DRO model with mixed-integer recourse. Such a model is difficult to solve directly using existing convex approximations~\cite{nemirovski2007convex,xie2022optimized} or conic reformulations~\cite{chen2024data}. RO avoids explicit distributional assumptions but depends on the uncertainty-set design. Box sets used for traffic demand and renewable energy output \cite{wei2014robust,wei2016robust} ignore spatio-temporal correlations and can produce overly conservative plans. Convex-hull sets preserve part of the empirical dependence structure \cite{velloso2019two}, but their complexity grows with dimension and they may contain unobserved combinations of uncertain quantities. Data-driven uncertainty sets retain the distribution-free nature of RO while using historical observations to capture spatio-temporal dependencies~\cite{mancilla2019power}, making them suitable for DWCL planning under correlated uncertainties. However, existing data-driven RO models mainly assess out-of-sample performance through empirical tests and lack risk quantification for planning decisions under unseen scenarios.

This paper proposes a data-driven robust DWCL planning framework with an MUE model and tunable profit--risk trade-off. The main contributions are summarized as follows:
\begin{itemize}
	\item A two-stage robust DWCL planning model is formulated for CTPNs to coordinate DWCL investment with CTPN operation. A data-driven uncertainty set is constructed from joint historical profiles of traffic demand and RES output to preserve their spatio-temporal correlations without imposing parametric distributional assumptions.
	
	\item A MUE model is developed for EV traffic assignment in TNs with DWCLs. Battery SoC dynamics and endogenous charging decisions are incorporated to capture EV routing and charging behavior more realistically.
	
	\item The proposed two-stage RO model with mixed-integer recourse is reformulated as an equivalent single-stage scenario-based problem and solved by a customized nested column-and-constraint generation (NCCG) algorithm.
	
	\item An \textit{a posteriori} feasibility certificate based on the number of support scenarios identified at optimality is developed to quantify out-of-sample risk. A risk-guided scenario-enrichment procedure is designed to expand the scenario set until the resulting risk bound meets a prescribed tolerance, enabling a tunable profit--risk trade-off.
\end{itemize}

The remainder of this paper is organized as follows. Section~II formulates the data-driven robust DWCL planning model with MUE-based traffic assignment. Section~III presents the solution algorithm and out-of-sample risk-quantification framework. Section~IV reports the case-study results, and Section~V concludes the paper.

\section{Data-Driven Two-Stage Robust DWCL Planning Model with Modified User Equilibrium}
\begin{figure}[!t]\vspace{0em}
	\centering
	\includegraphics[width=8.4cm]{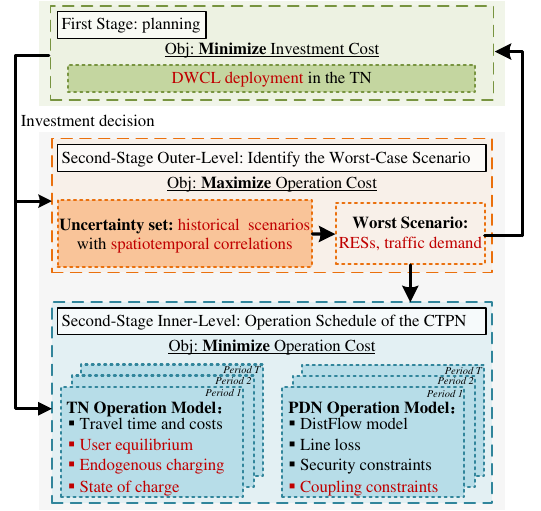}
	\caption{Proposed Data-Driven Robust DWCL Planning Framework with Modified User Equilibrium.}
	\label{fig2}
\end{figure}

This section presents the proposed data-driven two-stage robust DWCL planning model in the CTPN, as illustrated in Fig.~\ref{fig2}. The first stage determines the strategic investment decisions for DWCLs. The second stage is formulated as a bilevel max--min program. At the outer level, the worst-case realization of RES output and traffic demand is selected from a data-driven uncertainty set constructed from historical scenarios. At the inner level, the daily operational recourse in the CTPN is computed, with traffic assignment modeled by the proposed MUE incorporating SoC dynamics and endogenous charging decisions.
\subsection{Objective}
\subsubsection{First-Stage Investment Objective}
The first-stage objective is to minimize the annualized DWCL investment cost $f^{\rm inv}$ in (\ref{A_1}), where $\ell_{a}$ is the length of link $a$, $\mathcal{X}_{a}\in\{0,1\}$ indicates whether a DWCL is deployed on link $a$, and $\mathcal{C}_{\ell}$ is the unit-length DWCL investment cost. The capital recovery factor $\frac{\zeta(1+\zeta)^{\tau}}{(1+\zeta)^{\tau}-1}$ converts the one-time investment into an equivalent annual cost, where $\zeta$ is the discount rate and $\tau$ is the equipment lifetime \cite{wang2023robust}.
\begin{equation}\hspace{-6mm}
	\min \quad f^{\rm inv} 
	= \frac{\zeta(1+\zeta)^{\tau}}{(1+\zeta)^{\tau}-1}
	\sum_{a\in T_{A}} \ell_{a}\,\mathcal{X}_{a}\,\mathcal{C}_{\ell}
	\label{A_1}
\end{equation}

\subsubsection{Second-Stage Operational Recourse Objective}
The second stage is formulated as the $\max$--$\min$ problem in (\ref{B_1}). Following existing studies on CTPNs~\cite{wei2016optimal, lv2019optimal, wei2016robust, zhou2023planning}, an independent system operator is assumed to jointly manage the TN and the PDN. At the outer level, the worst-case uncertainty realization is selected to maximize the CTPN's daily operating cost. Conditional on this realization, the inner level minimizes the daily operating cost by optimizing recourse decisions, i.e., traffic assignment in the TN and power dispatch in the PDN. The daily operating cost is the sum of the TN cost and the PDN cost. Specifically, Eq.~(\ref{B_2}) defines the TN cost $F_{\rm TN}$ as the total vehicle travel cost, where $c_{a,t}$ is the travel cost on link $a$ in period $t$, and $x_{a,t}$ is the corresponding link flow. The PDN cost $F_{\rm PDN}$ in (\ref{B_3}) combines the cost of purchased electricity and the penalty for RES curtailment. Here, $p^{\rm g}_{j,t}$ and $p^{\rm d}_{j,t}$ denote purchased power and curtailed RES power at node $j$ in period $t$, respectively, and $a_{j,t}^{\rm g}$ and $a_{j,t}^{\rm d}$ are the corresponding unit-cost coefficients.
\begin{equation}
	\max\min \ F_{\rm TN} + F_{\rm PDN}
	\label{B_1}
\end{equation}
\begin{equation}\hspace{0mm}
	F_{\rm TN} = \sum_{t\in T_{T}}\sum_{a\in T_{A}} c_{a,t}\, x_{a,t}
	\label{B_2}
\end{equation}
\begin{equation}
	\begin{aligned}
		F_{\rm PDN} = \sum_{t\in T_{T}}\sum_{j\in E_{M}} \big(a_{j,t}^{\rm g}p^{\rm g}_{j,t} + a_{j,t}^{\rm d}p^{\rm d}_{j,t}\big)
	\end{aligned}
	\label{B_3}
\end{equation}

\subsection{Second-Stage Operation}
\subsubsection{Outer-Level Data-Driven Uncertainty Set}
The outer level of the second stage uses a data-driven uncertainty set $\bm{\Xi}$ constructed directly from joint historical profiles of RES output and traffic demand. Each element of $\bm{\Xi}$ represents one observed joint profile over the planning horizon. This construction preserves the observed spatio-temporal correlations without imposing parametric distributional assumptions. The scenario in this finite set that maximizes the CTPN's daily operating cost is selected as the worst-case realization.

\begin{figure}[!t]
	\centering
	\includegraphics[width=8cm]{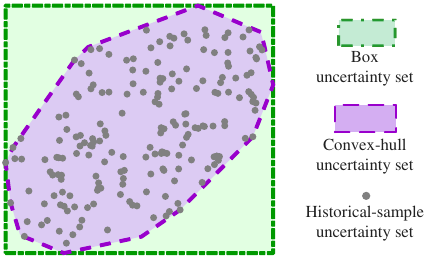}
	\caption{Comparison of Box, Convex-hull, and Proposed Uncertainty Sets.}
	\label{fig3}
\end{figure}

Fig.~\ref{fig3} compares the historical-sample set with two commonly used alternatives \cite{wei2016robust,velloso2019two}. The conventional box uncertainty set (green frame) \cite{wei2016robust} bounds each uncertain component independently and ignores spatio-temporal correlations in the data. Its worst-case realization can lie at an unrealistic corner of the box, leading to overly conservative decisions. The convex-hull set (purple boundary) \cite{velloso2019two} preserves part of the empirical dependence structure and is generally less conservative than the box set. However, its geometric complexity increases with dimension, and it may contain combinations of RES output and traffic demand that are not observed in the historical data. This can reduce the interpretability of the resulting worst-case scenarios.

By contrast, the historical-sample set (gray dots) contains only observed profiles, excluding unobserved combinations while preserving their empirical spatio-temporal correlations. It is used to identify worst-case scenarios and forms the basis for the scenario-approach reformulation and risk-quantification framework in Section~III.

\subsubsection{Inner-Level MUE-Based Traffic Assignment in the TN} 
Building on the classical UE model \cite{wei2016optimal,lv2019optimal,shi2022efficient}, an MUE model is developed for traffic assignment in the TN. The SoC along each path is tracked, and terminal-SoC requirements are imposed to ensure energy-feasible trips. Unlike the models in \cite{manshadi2017wireless,liu2024traffic}, which account for SoC dynamics but assume that EVs charge whenever they traverse a DWCL, the proposed MUE represents the charging decision on each DWCL segment with a binary variable. EVs with sufficient SoC may therefore traverse a DWCL without charging. The MUE is defined by constraints~(\ref{M_1})--(\ref{M_7}).

\textbf{a) Flow Conservation:} Eq.~(\ref{M_1}) ensures traffic flow conservation for each origin–destination (OD) pair $rs$, where the total traffic demand equals the sum of path flows.
\begin{equation}
	q_{t}^{rs} = \sum_{k\in T_{K}} f_{k,t}^{rs}, \ \forall rs, t
	\label{M_1} 
\end{equation}
where $q_{t}^{rs}$ is the demand of OD pair $rs$ in period $t$, and $f_{k,t}^{rs}$ is the corresponding flow on path $k$ connecting $rs$.

The link flow $x_{a,t}$ is derived from the path flows $f_{k,t}^{rs}$, as shown in (\ref{M_2}):
\begin{equation} \hspace{0mm}  x_{a,t} = \sum_{rs\in T_{W}}\sum_{k\in T_{K}}\delta_{ak}^{rs}f_{k,t}^{rs}, \ {\forall}a,t \label{M_2} 
\end{equation}
where $\delta_{ak}^{rs}$ is a binary incidence parameter, equal to 1 if link $a$ lies on path $k$
for OD pair $rs$, and 0 otherwise.

\textbf{b) Travel Cost Under Endogenous Charging:} The link travel time is modeled as a function of congestion using the Bureau of Public Roads function~\cite{wei2016optimal}:
\begin{equation}
	t_{a,t} = t_{a}^{0}\left[1+0.15\left({x_{a,t}}/{c_{a}^{0}}\right)^{4}\right], 
	\ \forall a, t
	\label{M_3} 
\end{equation}
where $t_{a}^{0}$ is the free-flow travel time on link $a$ and $c_{a}^{0}$ is the corresponding road capacity.

The path travel cost $c_{k,t}^{rs}$ is calculated as the sum of all link travel costs along the path, as defined in (\ref{M_4}) \cite{wei2016optimal}. The link travel cost includes the travel-time cost and charging cost, corresponding to the first and second terms in (\ref{M_4}), respectively.
\begin{equation}
	\begin{aligned}
		c_{k,t}^{rs} = &\sum_{a \in T_{A}} 
		(\omega t_{a,t} + \omega^{\rm c}\mu^{{\rm c},k}_{a,t}t_{a,t}e^{\rm c})
		\delta_{ak}^{rs}, \forall rs, k, t
	\end{aligned}
	\label{M_4} 
\end{equation}
where $\omega$ and $\omega^{\rm c}$ denote the monetary values of travel time and charging, respectively, the binary variable $\mu^{{\rm c},k}_{a,t}$ indicates whether charging occurs on link $a$ along path $k$ at time $t$, and $e^{\rm c}$ is the wireless charging power. Note that charging is permitted only on DWCLs (i.e., when $\mathcal{X}_{a}=1$), which is enforced by 
\begin{equation}
	\mu^{{\rm c},k}_{a,t} \leq \mathcal{X}_{a}, \ \forall k, a, t
	\label{M_5} 
\end{equation}

\textbf{c) User Equilibrium Under Endogenous Charging:} The self-interested route choice under endogenous charging is represented by the UE conditions in (\ref{M_6}). At equilibrium, each EV selects a minimum-cost path. All used paths for an OD pair have the same minimum cost $\pi_{t}^{rs}$, whereas every unused path has an equal or higher cost.
\begin{equation}\hspace{-0em}
	\begin{cases}
		f_{k,t}^{rs} \geq 0, c_{k,t}^{rs}-\pi_{t}^{rs} \geq 0, \forall rs, k, t\\[4pt]
		f_{k,t}^{rs}\big(c_{k,t}^{rs}-\pi_{t}^{rs}\big) = 0, \forall rs, k, t
	\end{cases}
	\!\!\!\!\!\!\!	\label{M_6}
\end{equation}

\textbf{d) State of Charge Under Endogenous Charging:} The SoC dynamics for an EV along each link are captured in (\ref{M_7}). The SoC $\mathcal{L}_{a^{\rm d},t}^{k,rs}$ at the downstream node $a^{\rm d}$ of link $a$ equals the upstream SoC $\mathcal{L}_{a^{\rm u},t}^{k,rs}$ minus normalized energy consumption on the link and plus normalized energy charged on the DWCL.
\begin{equation}				
	\begin{aligned}\hspace{-0em}
		\mathcal{L}_{a^{\rm d},t}^{k,rs} =&  \mathcal{L}_{a^{\rm u},t}^{k,rs} - \ell_{a}e/E + \mu^{{\rm c},k}_{a,t}t_{a,t}e^{\rm c}\alpha/E, \forall rs, a, k, t
	\end{aligned}
	\label{M_7} 
\end{equation}
where $\ell_{a}$ is the length of link $a$, $e$ is the energy consumption per kilometer, $E$ is the EV battery capacity, and $\alpha$ is the wireless power-transfer efficiency.

\subsubsection{Inner-Level Coupling Constraint} The coupling constraint between the TN and PDN is formulated in (\ref{M_8}), where the net wireless charging power $p^{\rm co}_{j,t}$ at bus $j$ in period $t$ is determined as follows:
\begin{equation}
	\begin{aligned}	  
		p^{\rm co}_{j,t} = 	&\sum_{rs\in T_{W}}\sum_{k\in T_{K}}\sum_{a\in T_{A}}\delta_{ak}^{rs}f_{k,t}^{rs}\mu^{{\rm c},k}_{a,t}e^{\rm c}, {\forall} j,t
	\end{aligned}
	\label{M_8}
\end{equation}

\subsubsection{Inner-Level Power Flow Optimization in the PDN} 
The PDN is modeled using the DistFlow formulation (\ref{N_1})–(\ref{N_6}) with line losses \cite{gan2021tri,wang2023joint}. 
\begin{align} 
	&\sum_{ij\in E_{\rm L}}(P_{ij,t}-r_{ij}\iota_{ij,t})+p^{\rm g}_{j,t}+(p^{\rm pv}_{j,t}-p^{\rm d}_{j,t})
	\notag\\
	&=\sum_{jk\in E_{\rm L}}P_{jk,t}+p_{j,t}^{\rm lo}+p_{j,t}^{\rm co},\ \forall j,t
	\label{N_1}\\
	&\sum_{ij\in E_{\rm L}}(Q_{ij,t}-x_{ij}\iota_{ij,t})+q^{\rm g}_{j,t}
	=\sum_{jk\in E_{\rm L}}Q_{jk,t}+q_{j,t}^{\rm lo},\ \forall j,t
	\label{N_2}\\
	&v_{i,t}-2(r_{ij}P_{ij,t}+ x_{ij}Q_{ij,t})+(r_{ij}^{2} + x_{ij}^{2})\iota_{ij,t}
	= v_{j,t},\ \forall ij,t
	\label{N_3}\\
	&\begin{Vmatrix} 2P_{ij,t} \\ 2Q_{ij,t}\\ \iota_{ij,t}-v_{i,t}\end{Vmatrix}_2
	\leq \iota_{ij,t}+v_{i,t}, \ \forall ij,t
	\label{N_4}\\
	&v_{j}^{{\rm min}} \leq {v}_{j,t} \leq v_{j}^{{\rm max}},\ 
	0 \leq \iota_{ij,t} \leq \iota_{ij}^{{\rm max}},\ \forall j,t,ij
	\label{N_5}\\
	&p^{{\rm min}}_{j} \leq p_{j,t}^{\rm g}\leq p^{{\rm max}}_{j},\ 
	q_{j}^{{\rm min}} \leq q_{j,t}^{\rm g}\leq q_{j}^{{\rm max}},\ \forall j,t
	\label{N_6}
\end{align}

Constraints~(\ref{N_1}) and (\ref{N_2}) represent nodal active and reactive power balance, respectively. Constraint~(\ref{N_3}) represents the voltage drop along the distribution lines. Constraint~(\ref{N_4}) describes the relationship among power, voltage, and current in a second-order cone (SOC) form \cite{wei2016optimal}. Constraint~(\ref{N_5}) restricts node voltages and line currents within operational limits. Constraint~(\ref{N_6}) bounds active and reactive power generation. Here, $r_{ij}$ and $x_{ij}$ are the resistance and reactance of line $ij$; $P_{ij,t}$ and $Q_{ij,t}$ are the active and reactive power flows on line $ij$ in period $t$; $\iota_{ij,t}$ is the squared current magnitude of line $ij$; $v_{i,t}$ is the squared voltage magnitude at node $i$; $p^{\rm g}_{j,t}$ and $q^{\rm g}_{j,t}$ are the active and reactive generation; $p^{\rm lo}_{j,t}$ and $q^{\rm lo}_{j,t}$ are the active and reactive loads; $p^{\rm pv}_{j,t}$ and $p^{\rm d}_{j,t}$ are the PV generation and curtailed PV power; and $v_{j}^{\min}, v_{j}^{\max}, p_{j}^{\min}, p_{j}^{\max}, q_{j}^{\min}, q_{j}^{\max}, \iota_{ij}^{\max}$ denote the corresponding bounds.

\section{Solution Method and Tunable Risk Quantification} 

\begin{figure}[!t]\vspace{-0em}
	\centering
	\includegraphics[width=8.4cm]{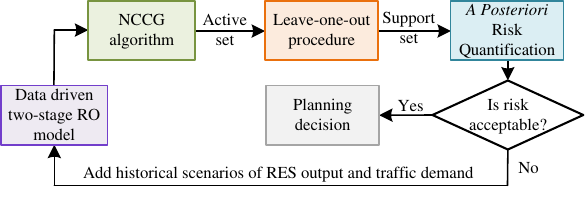}
	\caption{Proposed Solution and Profit--Risk Tuning Framework.}
	\label{fig4}
\end{figure}

This section presents the model linearization, the equivalent single-stage scenario-approach (SA) reformulation, and the customized NCCG algorithm. A leave-one-out procedure is applied to the active scenario set to identify a support set, whose cardinality is used to evaluate an \textit{a posteriori} feasibility certificate for out-of-sample risk. If the resulting risk bound exceeds a prescribed tolerance, the scenario set is enriched and the model is re-solved. This risk-guided procedure enables a tunable profit--risk trade-off and is illustrated in Fig.~\ref{fig4}.

\subsection{Model Linearization}
\subsubsection{Travel-Cost Reformulation}
The term $c_{a,t}x_{a,t}$ in (\ref{B_2}) is nonlinear. Using (\ref{M_1}), (\ref{M_2}), and the UE conditions in (\ref{M_6}), the TN cost can be rewritten as \cite{wei2016optimal}.
\begin{equation}
	\hspace{-2mm}
	\begin{aligned}
		&\sum_{t\in T_{T}}\sum_{a\in T_{A}} c_{a,t}x_{a,t}
		=\sum_{t\in T_{T}}\sum_{rs\in T_{W}}\sum_{k\in T_{K}}\sum_{a\in T_{A}} c_{a,t}\,f^{rs}_{k,t}\delta_{ak}^{rs}\\
		&=\sum_{t\in T_{T}}\sum_{rs\in T_{W}}\sum_{k\in T_{K}} c_{k,t}^{rs}f_{k,t}^{rs}=\sum_{t\in T_{T}}\sum_{rs\in T_{W}}\pi_{t}^{rs}q_{t}^{rs}
	\end{aligned}
	\label{H_0}
\end{equation}
\subsubsection{Piecewise Linearization}
Equation~(\ref{M_3}) is nonlinear because of the quartic term $\left(x_{a,t}/c_{a}^{0}\right)^{4}$. Following \cite{gan2021tri}, this term is approximated by a piecewise-linear function:
\begin{equation}
	t_{a,t} = t_{a}^{0} + {0.15\,t_{a}^{0}}/{(c_{a}^{0})^{4}}\,\widetilde{x}_{a,t}, \quad \forall a,t,
	\label{H_1}
\end{equation}\vspace{0em}
\begin{align}
	\left\{
	\begin{aligned}
		&\widetilde{x}_{a,t} = \sum_{\tau} x_{a,t,\tau}\,\upsilon_{a,t,\tau}, \quad x_{a,t} = \sum_{\tau} x_{a,t,\tau}, \\
		&0 \leq x_{a,t,\tau} \leq \Delta x_{a,t,\tau}, \quad \forall \tau
	\end{aligned}
	\right.
	\label{H_2}
\end{align}
where $\widetilde{x}_{a,t}$ is the piecewise-linear approximation of $x_{a,t}^{4}$. The parameter $\Delta x_{a,t,\tau}$ denotes the width of segment $\tau$, $x_{a,t,\tau}$ is the portion of $x_{a,t}$ assigned to this segment, and $\upsilon_{a,t,\tau}$ is the corresponding slope.
\subsubsection{Linearization Using the Big-M Method}
The bilinear term $\mu^{\rm c,k}_{a,t} t_{a,t}$ in (\ref{M_4}) is linearized via the big-$M$ method \cite{lv2019optimal}. By introducing an auxiliary variable $w^{k}_{a,t}$ and a sufficiently large constant $M$, the product $\mu^{\rm c,k}_{a,t} t_{a,t}$ is replaced by 
\begin{align}
	\left\{
	\begin{aligned}
		& 0 \leq t_{a,t} - w^{k}_{a,t} \leq M(1 - \mu^{\rm c,k}_{a,t}) \\
		& 0 \leq w^{k}_{a,t} \leq M \mu^{\rm c,k}_{a,t}
	\end{aligned}
	\right.
	\label{H_3}
\end{align}
Similarly, the complementarity condition in (\ref{M_6}) is linearized by the big-$M$ method.

\subsection{Reformulation From Two-Stage RO to Single-Stage SA}
After linearization, the proposed two-stage data-driven robust DWCL planning problem can be represented in the following compact form:
\begin{subequations}
	\begin{numcases}{}
		\min_{{\bm x}\in {\bm X}}~\bm{c}^{\top}\bm{x}
		+ \max_{\bm \xi\in \bm \Xi}\min_{\bm y \in {\bm Y},\,\bm z \in {\bm Z}} \bm{d}^{\top}\bm{y} \label{I_1}\\
		\text{s.t.}\ \bm{A}\bm{x}\leq \bm{g},\ \bm{x}\in \{0,1\} \label{I_2}\\
		\bm{B}\bm{x}+\bm{C}\bm{\xi} + \bm{D}\bm{y} + \bm{E}\bm{z} \leq \bm{h},\ \bm{z}\in \{0,1\}
		\label{I_3}
	\end{numcases}\label{I}
\end{subequations}
\par\noindent
In \eqref{I}, vector $\bm{x}$ collects the first-stage investment decisions, while $\bm{\xi}$ characterizes the second-stage outer-level uncertainty. Vectors $\bm{y}$ and $\bm{z}$ denote the continuous and binary recourse variables at the inner level of the second stage, respectively. Matrices $\bm{A}-\bm{E}$ contain the corresponding coefficients, and $\bm{g}, \bm{h}$ are constant vectors. Constraints \eqref{I_2} and \eqref{I_3} specify the first- and second-stage constraints, respectively.

Problem~\eqref{I} enforces feasibility for all scenarios in $\bm\Xi$ but does not quantify the violation probability under unseen scenarios. Because $\bm\Xi$ is finite, \eqref{I} can be equivalently written as the following single-stage SA model:
\begin{align}
	\left\{
	\begin{aligned}
		&\min_{\bm x\in\bm X,\ \eta\in\mathbb{R}}\ 
		\bm c^{\top}\bm x+\eta \\
		&\text{s.t. }\ \eta \ge \bm d^{\top}\bm y^{(s)},\ s=1,\ldots,S \\
		&\bm A\bm x\le \bm g, \ \bm B\bm x+\bm C\bm \xi^{(s)}+\bm D\bm y^{(s)}+\bm E\bm z^{(s)}\le \bm h\\
	\end{aligned}\label{J}
	\right.
\end{align}
Here, $\eta$ is an epigraph variable that upper-bounds the worst-case second-stage cost. The index $s\in\{1,\ldots,S\}$ denotes a scenario in $\bm\Xi=\{\bm\xi^{(s)}\}_{s=1}^{S}$, where $S\coloneqq|\bm\Xi|$ is the number of scenarios. The vectors $(\bm y^{(s)},\bm z^{(s)})$ are scenario-specific copies of the continuous and binary recourse variables. The equivalence between \eqref{I} and \eqref{J} is proved in Appendix~A.

Recasting the two-stage RO problem \eqref{I} in the SA form \eqref{J} is mainly motivated by its suitability for finite-sample data-driven analysis \cite{garatti2022risk}. In particular, the optimal solution of \eqref{J} is supported only by a subset of the sampled scenarios, referred to as the support scenarios. This structural property allows us to directly apply scenario-optimization results to derive statistical feasibility guarantees for unseen scenarios, expressed in terms of violation probabilities. These guarantees are \emph{a posteriori}: they are evaluated after solving the SA problem and identifying the support scenarios, and they do not require any parametric assumption on the underlying probability distribution. \vspace{-1em}

\subsection{NCCG Algorithm for Solving the SA Model}
Since the number of scenarios $S$ is often large, directly solving the SA problem \eqref{J} over all scenarios is computationally inefficient and may even be infeasible. Building on the ideas in \cite{long2012exact}, we therefore adopt a customized NCCG algorithm. At the outer level, a CCG loop iteratively updates a restricted scenario subset $\bm\Omega\subseteq\bm\Xi$ (the active set) and solves a master problem that yields a lower bound on the optimal objective. For each candidate first-stage decision produced by this master problem, an inner CCG loop is invoked to tackle the mixed-integer recourse problem and to compute both a worst-case scenario and an upper bound on the objective value. The worst-case scenario identified by the inner loop is then appended to the active set ${\bm\Omega}$, and the outer loop proceeds until the resulting optimality gap falls below a prescribed tolerance. The overall procedure is detailed as follows.

\subsubsection{Outer-level master problem (OMP)}
Given a current active set ${\bm\Omega}=\{s_{1},s_{2},...,s_{R}\}\subseteq\bm\Xi$, the OMP in \eqref{K} is a relaxation of \eqref{J} that enforces the SA constraints only for scenarios in ${\bm\Omega}$. At outer iteration $r$, solving \eqref{K} yields a candidate first-stage decision $\bm x^{*}$, an epigraph value $\eta$, and a lower bound $\mathrm{LB}^{*} = \bm c^{\top}\bm x^{*}+\eta$ \cite{zeng2013solving}. The incumbent outer lower bound is updated as $\mathrm{LB} \leftarrow \max\{\mathrm{LB},\mathrm{LB}^{*}\}$. The decision $\bm x^{*}$ is then passed to the inner-level CCG to evaluate the associated worst-case recourse cost.
\begin{align}
	\left\{
	\begin{aligned}
		&\min_{\bm{x}\in\bm X,\,\eta\in\mathbb{R}} \ 
		\bm{c}^{\top}\bm{x}+\eta \\
		&\text{s.t.}\ \eta \ge \bm{d}^{\top}\bm{y}^{(s)}, \ \forall 1 \leq s \leq \mathrm{R}\\ 
		&\bm{A}\bm{x}\le \bm{g}\\
		&\bm{B}\bm{x}+\bm{C}\bm{\xi}^{(s)}+\bm{D}\bm{y}^{(s)}+\bm{E}\bm{z}^{(s)}\le \bm{h}, \ \forall 1 \leq s \leq \mathrm{R}
	\end{aligned}\label{K}
	\right.
\end{align}
\subsubsection{Inner-level CCG} For the fixed first-stage decision $\bm x^{*}$, the second-stage recourse problem with integer variables is solved by an inner-level CCG loop. This loop alternates between an inner-level master problem (IMP) and an inner-level subproblem (ISP).

\textit{a) IMP:} The IMP in \eqref{L} is obtained by dualizing the ISP with fixed integer recourse variables \cite{long2012exact}. At inner iteration $l$, given the first-stage decision $\bm x^{*}$ and the current set of mixed-integer recourse solutions $\bm Z$, the IMP searches for the worst-case scenario $\bm\xi^{\star}$ over the uncertainty set $\bm\Xi$ and yields an inner upper bound $\mathrm{ub}^{*} = \bm c^{\top}\bm x^{*} + \theta$.
\begin{subequations}
	\begin{numcases}{} 
		\max_{\bm\xi\in \bm \Xi, \bm\lambda,\theta} \theta\\			
		{\rm s.t.} \ \theta\leq (\bm\lambda^{\top}_{q}\bm{B}\bm{x}^{*}+\bm\lambda^{\top}_{q}\bm{C}\bm{\xi} + \bm\lambda^{\top}_{q}\bm{E}\bm{z}^{q*} 		\label{HW_1}\\
		\quad \ - \bm\lambda^{\top}_{q}\bm h) 
		,\ \forall 1 \leq q \leq \mathrm{R}	\label{HW_2}\\
		\hspace{6mm}\bm{d}^{\top} + \bm\lambda^{\top}_{q}\bm D \leq 0,\ \forall 1 \leq q \leq \mathrm{R}		
	\end{numcases}
	\label{L}
\end{subequations}

\textit{b) ISP:} For the fixed pair $(\bm x^{*},\bm\xi^{\star})$, the ISP in \eqref{M} \cite{zeng2013solving} evaluates the recourse cost by solving
\begin{align}
	\left\{
	\begin{aligned}
		&\min_{\bm{y}\in \bm Y,\,\bm{z}\in \bm Z}\  \bm{d}^{\top}\bm{y} \\
		&\text{s.t.}\ \bm{B}\bm{x}^{*}+\bm{C}\bm{\xi}^{*}+\bm{D}\bm{y}+\bm{E}\bm{z}\le \bm{h}
	\end{aligned}
	\right.
	\label{M}
\end{align}
which yields optimal continuous recourse variables $\bm y^{\star}$, integer recourse variables $\bm z^{\star}$, and an inner lower bound $\mathrm{lb}^{*} = \bm c^{\top}\bm x^{*} + \bm d^{\top}\bm y^{\star}$. The variables $\bm z^{\star}$ are then added to the set $\bm Z$ used in the IMP. The inner CCG iterations continue until the gap $(\mathrm{ub}^{*} - \mathrm{lb}^{*})/\mathrm{ub}^{*}$ falls below a prescribed tolerance.

\subsubsection{Convergence and Scenario Update} 
After the inner CCG loop converges for $\bm x^{*}$, the worst-case scenario $\bm\xi^{\star}$ and the corresponding upper bound $\mathrm{ub}^{*}$ are returned to the outer loop, and $\mathrm{UB}$ is updated accordingly. If $(\mathrm{UB}-\mathrm{LB})/\mathrm{UB}\leq\varepsilon$, the incumbent planning decision is returned as an $\varepsilon$-optimal solution. Otherwise, the active set is augmented as $\bm\Omega\leftarrow\bm\Omega\cup\{\bm\xi^{\star}\}$, and the OMP in \eqref{K} is solved again. Steps~1.1--1.4 of Algorithm~1 summarize this procedure.

\begin{table}[!t]\vspace{0.5em}
	\renewcommand\arraystretch{1.1}
	\setlength{\tabcolsep}{0.02mm}
	\centering
	\begin{tabular}{l}
		\toprule[2pt]
		\textbf{Algorithm 1:} \text{NCCG Algorithm with Tunable Profit--Risk Trade-off} \\
		\midrule[1pt]
		\textbf{0: Initialization} \\
		\qquad Select an initial scenario set $\bm\Xi$, set the active scenario set ${\bm\Omega}\!\gets\!\varnothing$,\\
		\qquad  set optimization tolerance $\varepsilon$ and risk tolerance $\bar{\varepsilon}$, outer iteration \\
		\qquad  $r\!\gets\!0$, and outer bounds $\mathrm{LB}\!\gets\!-\infty$, $\mathrm{UB}\!\gets\!+\infty$.\\
		\addlinespace[2pt]
		\textbf{1: NCCG Algorithm} \\
		1.1~\textbf{Outer-level master problem (OMP):} For the current active scenario \\
		\qquad set ${\bm\Omega}$, solve the OMP (\ref{K}) to obtain the planning decision $\bm x^{*}$,  \\
		\qquad mixed-integer recourse variable $\bm z^{*}$ and epigraph variable $\eta^{*}$; \\
		\qquad update $\mathrm{LB}\!\leftarrow\!\max\{\mathrm{LB},\,\bm c^{\top}\bm x^{*}+\eta^{*}\}$.\\
		\addlinespace[2pt]
		1.2~\textbf{Inner-level CCG initialization:} \\
		\qquad Set inner bounds $\mathrm{lb}\!\gets\!-\infty$, $\mathrm{ub}\!\gets\!+\infty$, and inner iteration $l\!\gets\!0$. \\
		\addlinespace[2pt]
		1.3~\textbf{Inner-level CCG iterations:} \\
		\qquad\quad 1.3.1~\textbf{Inner master problem (IMP):} For fixed $\bm x^{*}$ and $\bm z^{*}$,  \\
		\qquad\quad\quad solve the IMP (\ref{L}) to obtain a worst-case scenario $\bm\xi^{*}$ from $\bm\Xi$ \\
		\qquad\quad\quad and update the inner bound $\mathrm{ub}\!\leftarrow\!\min\{\mathrm{ub},\,\bm c^{\top}\bm x^{*}+\theta^{*}\}$. \\
		\qquad\quad 1.3.2~\textbf{Inner subproblem (ISP):} For fixed $\bm x^{*}$ and $\bm\xi^{*}$, solve the \\
		\qquad\quad\quad ISP (\ref{M}) to obtain continuous recourse $\bm y^{*}$ and integer recourse \\
		\qquad\quad\quad $\bm z^{l}$; update the inner bound $\mathrm{lb}\!\leftarrow\!\max\{\mathrm{lb},\,\bm c^{\top}\bm x^{*}+\bm d^{\top}\bm y^{*}\}$.\\
		\qquad\quad 1.3.3~\textbf{Inner-level CCG convergence test:} If $({\mathrm{ub}-\mathrm{lb}})/{\mathrm{ub}} \le \varepsilon$, \\
		\qquad\quad\quad set $\mathrm{UB} \leftarrow \min\{\mathrm{UB},\,\mathrm{ub}\}$, exit inner CCG, and go to Step~1.4;   \\
		\qquad\quad\quad otherwise enlarge $\bm z^{*}\!\leftarrow\!\bm z^{*} \cup  \{\bm z^{l}\}$, $l\!\gets\!l+1$, and continue  \\
		\qquad\quad\quad  with Step~1.3.1.\\
		\addlinespace[2pt]
		1.4~\textbf{Outer-level CCG convergence test} \\
		\qquad If $({\mathrm{UB}-\mathrm{LB}})/{\mathrm{UB}} \le \varepsilon$, go to Step~2; otherwise set $r\!\gets\!r+1$, \\
		\qquad enlarge the active scenario set ${\bm\Omega}\!\leftarrow\!{\bm\Omega}\cup\{\bm \xi^{*}\}$, and return to Step~1.1. \\[1pt]
		\addlinespace[2pt]
		\textbf{2: Support-set extraction (parallel)} \\
		\qquad For each $k\in{\bm\Omega}$, resolve the outer OMP with active set ${\bm\Omega}\setminus\{k\}$; if  \\[1pt]
		\qquad the optimal objective strictly decreases, mark $k$ as supporting;  \\
		\qquad otherwise mark it as non-supporting. Set ${\bm\Omega}_{\mathrm{sup}}=\{k:\text{supporting}\}$  \\ 
		\qquad and $\kappa\coloneqq|{\bm\Omega}_{\mathrm{sup}}|$. \\
		\addlinespace[2pt]
		\textbf{3: \textit{A posteriori} risk quantification} \\
		\qquad Using $\kappa$ and $S\coloneqq|\bm\Xi|$, apply \textit{a posteriori} wait-and-judge bound  \\[1pt]
		\qquad (\ref{O_2})–(\ref{O_3}) to compute a feasibility guarantee, i.e., the  \\[1pt]
		\qquad violation level $\hat{\varepsilon}$ at confidence $1-\beta$ for $\bm x^\star$. \\
		\addlinespace[2pt]
		\textbf{4: Risk-acceptance test and scenario augmentation} \\
		\qquad If $\hat{\varepsilon} \le \bar{\varepsilon}$, terminate and go to Step~5; \\ 
		\qquad otherwise, augment the scenario set by adding $N_{\text{new}}$ additional \\[1pt]
		\qquad scenarios to $\bm\Xi$, i.e., $\bm\Xi \leftarrow \bm\Xi \cup \{\tilde{\bm\xi}^{(s)}\}_{s=1}^{N_{\text{new}}}$ and return to Step~0.\\
		\addlinespace[2pt]
		\textbf{5:} Output the investment decision $\bm x^\star$ and \textit{a posteriori} feasibility \\[1pt]
		\qquad  guarantee $\hat{\varepsilon}$ at confidence $1-\beta$. \\
		\bottomrule[2pt]
	\end{tabular}\vspace{0em}
\end{table}
\subsection{Support Scenario Extraction and Risk Quantification}
\subsubsection{Support Scenario Extraction}
The NCCG algorithm returns an active scenario set $\bm\Omega$ that reproduces the selected optimal solution of the full SA problem. Because $\bm\Omega$ is generated iteratively, it may contain scenarios that are redundant at optimality. Therefore, a support set $\bm\Omega_{\mathrm{sup}}\subseteq\bm\Omega$ is  identified using a sequential leave-one-out procedure for the subsequent risk analysis \cite{garatti2022risk,yan2022two}.

Starting from $\bm\Omega$, each scenario in the current set is removed in turn, and the restricted SA problem is re-solved. If the selected optimal solution remains unchanged, the scenario is discarded; otherwise, it is retained. The resulting support set has cardinality $\kappa\coloneqq|\bm\Omega_{\mathrm{sup}}|$. Its cardinality can be used as an upper bound on the scenario complexity, yielding a potentially conservative \textit{a posteriori} risk certificate \cite{garatti2025non}.

\subsubsection{Tunable Risk Quantification}
Given the final investment decision $\bm x^\star$, its violation probability is defined as \cite{garatti2025non}
\begin{equation}
	V(\bm x^\star) = \mathbb{P}_{\bm \varrho}\{\bm{d}^{\top}\bm{y}(\bm x^\star,\bm \varrho) > \eta^\star\},
	\label{O_1}
\end{equation}
which is the probability that the optimal recourse cost under an unseen scenario exceeds the epigraph value $\eta^\star$.

According to \textbf{Theorem~4} in \cite{garatti2025non}, for the nonconvex SA model \eqref{J}, a distribution-free \emph{a posteriori} feasibility guarantee holds at any confidence level $1-\beta\in(0,1)$:
\begin{equation}
	\mathbb{P}^{S}\left\{\, V(\bm x^\star) > \varepsilon(\kappa)\,\right\}\ \le\ \beta,
	\label{O_2}
\end{equation}
where $\mathbb{P}^{S}$ denotes probability with respect to the draw of the $S$ training scenarios, $\varepsilon(\kappa)\in[0,1]$ is a violation-probability bound, and $\kappa$ is the cardinality of the support set (number of support scenarios). Eq.~\eqref{O_2} states that, with confidence level $1-\beta$, the out-of-sample violation probability of $\bm x^\star$ under the true distribution $P$ does not exceed $\varepsilon(\kappa)$.
\begin{figure*}[!t]\vspace{0em}
	\centering
	\includegraphics[width=16cm]{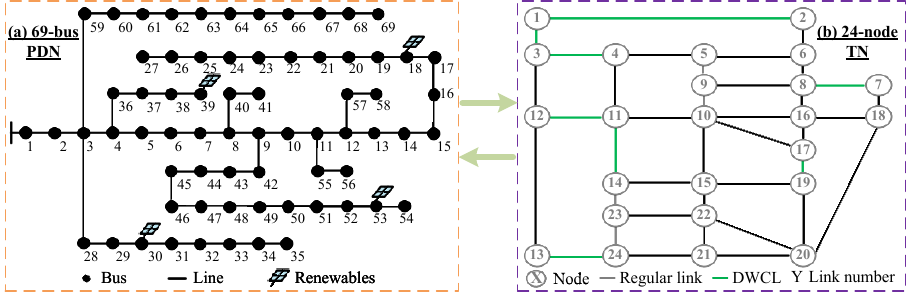}
	\caption{Topology of the coupled traffic--power network: (a) 69-bus distribution network and (b) 24-node Sioux Falls traffic network, where the green links denote the DWCLs deployed in Case V.}
	\label{fig5}\vspace{0em}
\end{figure*}

Given $\kappa$, the bound $\varepsilon(\kappa)$ is specified as the unique root in $(0,1)$ of the following polynomial equation \cite{garatti2025non}:
\begin{equation}\vspace{-0.3em}
	\frac{\beta}{S}\sum_{m=\kappa}^{S-1}\binom{m}{\kappa} \big(1-\varepsilon(\kappa)\big)^{\,m-\kappa} 
	- \binom{S}{\kappa} \big(1-\varepsilon(\kappa)\big)^{\,S-\kappa} = 0,
	\label{O_3}
\end{equation}
which can be computed reliably by a bisection search.
The bound $\varepsilon(\kappa)$ in \eqref{O_2}--\eqref{O_3} thus provides \textit{a posteriori} quantitative measure of the out-of-sample violation probability.

To make the risk level tunable, we introduce a user-specified risk tolerance $\bar{\varepsilon}$ and embed the support-set extraction steps into an outer loop that iteratively augments the scenario set. Let $\hat{\varepsilon} := \varepsilon(\kappa)$ denote the current risk bound. If $\hat{\varepsilon}$ exceeds $\bar{\varepsilon}$, additional scenarios are sampled and added to the scenario set $\bm\Xi$, and the NCCG procedure is re-run on the enlarged scenario set. This process is repeated until the resulting decision $\bm x^\star$ satisfies $\hat{\varepsilon}\le \bar{\varepsilon}$. The complete solution algorithm with tunable profit--risk trade-off is summarized in Algorithm~1.

\section{Case Studies}
This section assesses the performance of the proposed model on a CTPN consisting of a modified IEEE 69-node PDN and a 24-node Sioux Falls TN. All simulations are executed on a laptop with an Intel Core i7-11800H CPU and 16~GB of RAM, and all optimization models are solved using Gurobi.

\begin{table}[!t]
	\vspace{-0em}
	\renewcommand{\arraystretch}{1.1}
	\centering
	\caption{Planning and operating cost comparison under different traffic assignment models}
	\label{tab1}
	\setlength{\tabcolsep}{0.4mm}
	\begin{tabular}{c|c|ccc|c}
		\toprule[0.8pt]
		\multirow{2}{*}{\textbf{Case}} 
		& \multicolumn{1}{c|}{\textbf{Investment cost ($\$10^{6}$)}} 
		& \multicolumn{3}{c|}{\textbf{Operating cost ($\$10^{6}$)}} 
		& \multirow{2}{*}{\textbf{Total cost \textbf{($\$10^{6}$)} }}\\
		& {DWCL} & {TN} & {PDN} & {Penalty}\\
		\hline
		\textbf{I}   & 6.84 &3.65  &5.14   &1.13  & 16.76  \\
		\textbf{II}  & 7.46 &3.72  &4.62   &0     & 15.80\\
		\textbf{III} & 4.60 & 3.70 &4.54   & 0    & 12.84  \\
		\bottomrule[0.8pt]
	\end{tabular}	\vspace{0em}
\end{table}

\begin{figure}[!t]\vspace{0em}
	\centering
	\includegraphics[width=8.4cm]{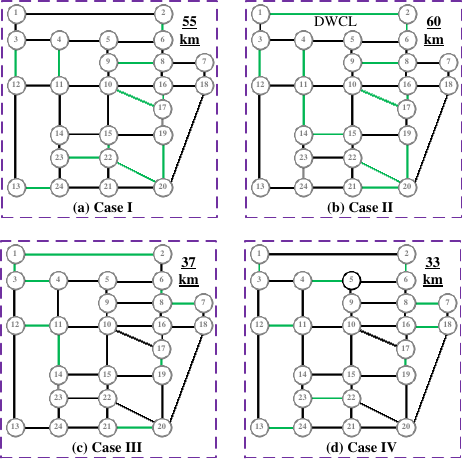}
	\caption{DWCL deployment results for different cases.}
	\label{fig6}
\end{figure}

\subsection{Simulation Settings}
Fig.~\ref{fig5}(a) shows the PDN topology. The power base is $S_{\rm base}=100$~MVA, and the voltage limits are set to $v_{j}^{\rm min}=0.83$ and $v_{j}^{\rm max}=1.10$ (p.u.) \cite{wei2016optimal}. Other PDN parameters follow \cite{cui2025modeling}. Fig.~\ref{fig5}(b) shows the TN topology. The base traffic demand is 100~veh/h \cite{wei2016optimal}. Link capacity, free-flow travel time, and link length follow \cite{ukkusuri2009methodology}. The wireless charging power is set to 60~kW \cite{liu2024traffic}. Each EV has a 100~kWh battery and an energy consumption rate of 0.185~kWh/km \cite{liu2024traffic}. The value of time and charging price are set to 11.26~\$/h and 1/15~\$/kWh, respectively \cite{cui2026ev}. The unit investment cost of DWCL is $\$1.55\times 10^{6}$/km \cite{liu2024traffic, xia2021distributed}. The facility lifetime is 20~years. The confidence parameter is set to $\beta=0.05$ (95\% confidence), and the risk tolerance is set to $\bar{\varepsilon}=0.01$. Following the proposed construction, the uncertainty set is a purely data-driven historical-sample set $\bm\Xi$ (2000 scenarios), where each scenario is an observed joint spatio-temporal profile of PV output and traffic demand over 24~hours. The scenarios are randomly split into an in-sample set $\bm\Xi_{\rm in}$ (1000 scenarios) for solving all planning models and an out-of-sample set $\bm\Xi_{\rm out}$ (1000 scenarios) for performance assessment. 

Five cases for DWCL planning are designed:

$\bullet$ \textbf{Case I}: A conventional RO framework is used with a SoC-aware system-optimum traffic assignment model, where EVs charge whenever a DWCL is traversed.

$\bullet$ \textbf{Case II}: A conventional RO framework is used with a SoC-aware UE model, where EVs charge whenever a DWCL is traversed.

$\bullet$ \textbf{Case III}: A conventional RO framework is used with a SoC-aware UE model with endogenous charging decisions, where EVs may skip charging when their SoC is sufficient.

$\bullet$ \textbf{Case IV}: Unlike Case III, a conventional two-stage SP framework is used.

$\bullet$ \textbf{Case V}: Unlike Case III, the proposed data-driven two-stage RO framework with a tunable profit--risk trade-off is used.

Note that, for \textbf{Case~V}, the in-sample scenarios are not used all at once. Instead, the proposed risk-feasibility check starts from an initial subset of 200 scenarios and expands the subset by 200 scenarios per iteration until the prescribed risk tolerance is satisfied.

\begin{figure}[!t]\vspace{0em}
	\centering
	\includegraphics[width=9cm]{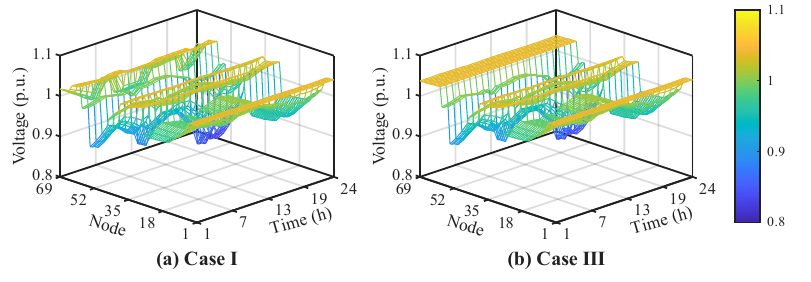}
	\vspace{-18pt}
	\caption{PDN voltage profiles: (a) \textbf{Case~I} and (b) \textbf{Case~III}.}
	\label{fig7}
\end{figure}

\subsection{Impact of Different Traffic Assignment Models}
Table~\ref{tab1} summarizes the annualized investment and operating costs for \textbf{Case~I--Case~III}. To ensure a fair comparison, the DWCL plans in \textbf{Case~I} and \textbf{Case~II} are first obtained using their respective traffic assignment models. These two plans are then re-evaluated under the same worst-case uncertainty scenario and the same traffic assignment model adopted in \textbf{Case~III}, so that all operating costs reflect a consistent representation of actual operations. Fig.~\ref{fig6} compares the resulting DWCL deployment locations across cases.

\textbf{Case~I} and \textbf{Case~II} are compared to analyze the effect of incorporating UE in traffic assignment. As shown in Fig.~\ref{fig6}, \textbf{Case~I} deploys 11 DWCLs, including DWCLs 2--6, 3--12, 4--11, 9--8, 10--17, 16--17, 15--22, 23--22, 19--20, 22--20, and 13--24. The total DWCL length is 55 km, and the annualized investment cost is $\$6.84\times 10^{6}$, as reported in Table~\ref{tab1}. In \textbf{Case~II}, the DWCL locations change noticeably, and the total DWCL length increases to 60 km, leading to an about 9.1\% higher annualized investment cost. This difference stems from the flow-assignment pattern: \textbf{Case~I} allocates traffic to minimize the system-wide cost, whereas \textbf{Case~II} follows self-interested UE routing, resulting in different traffic and charging-demand distributions.

\textbf{Case~II} and \textbf{Case~III} are further compared to assess the impact of endogenous charging under UE. Fig.~\ref{fig6} shows that the deployed DWCL length in \textbf{Case~III} drops to 37 km, which is 23 km shorter than that in \textbf{Case~II}. The corresponding annualized investment cost in \textbf{Case~III} is about 38.3\% lower than that in \textbf{Case~II}, as shown in Table~\ref{tab1}. The key reason is that endogenous charging does not force EVs to charge whenever a DWCL is traversed. As a result, there is no need to over-build DWCLs to support redundant charging for high-SoC users.

From an operational perspective under the endogenous-charging UE model, the plan obtained in \textbf{Case~I} leads to severe congestion in the PDN. Specifically, Fig.~\ref{fig7} shows voltage violations during time periods 19 and 20. As reported in Table~\ref{tab1}, this results in the highest total annualized cost among the three cases, reaching $\$16.76\times 10^{6}$. By contrast, \textbf{Case~II} avoids voltage violations, but its operating cost is about $\$1\times 10^{5}$ higher than that of \textbf{Case~III}, and its total annualized cost exceeds that of \textbf{Case~III} by more than 18.7\%. These results suggest that DWCL plans derived from \textbf{Case~I} and \textbf{Case~II} do not perform well under the actual operating setting represented by \textbf{Case~III}. With endogenous charging, EVs may skip charging on some DWCLs, so the realized charging demand can differ from what \textbf{Case~I} and \textbf{Case~II} implicitly assume, which in turn shifts the charging load distribution in the PDN.

\subsection{Impact of Different Uncertainty-Handling Methods}
This subsection compares \textbf{Case~III} (conventional RO), \textbf{Case~IV} (SP), and \textbf{Case~V} (the proposed data-driven RO model). Both in-sample and out-of-sample results are reported to demonstrate the practical benefits of the proposed model.

\subsubsection{In-sample Performance Comparison} 
\begin{figure}[!t]	\vspace{0em}
	\centering
	\includegraphics[width=9cm]{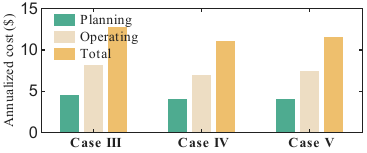}
	\vspace{-18pt}
	\caption{In-sample annualized investment and operating costs for \textbf{Case~III}--\textbf{Case~V}.}
	\label{fig8}	\vspace{-0em}
\end{figure}

The DWCL deployment results of \textbf{Case~III}, \textbf{Case~IV}, and \textbf{Case~V} are shown in Fig.~\ref{fig6}(c), Fig.~\ref{fig6}(d), and Fig.~\ref{fig5}(b), respectively. Across the three cases, both the deployment locations and the overall coverage differ. Fig.~\ref{fig8} shows the annualized investment and operating costs for \textbf{Case~III--Case~V}. \textbf{Case~III} yields the highest total annualized cost, reaching  $\$12.84\times 10^{6}$, whereas \textbf{Case~IV} achieves the lowest total annualized cost, around $\$11.13\times 10^{6}$. The total annualized cost of \textbf{Case~V} falls between the two. These differences mainly come from how uncertainty is modeled. In \textbf{Case~III}, the uncertainty set is built in a worst-case manner, which typically leads to conservative DWCL plans and higher costs. In contrast, \textbf{Case~IV} uses a finite set of scenarios to describe uncertainty, which often matches the in-sample data more closely and therefore yields lower in-sample costs. In \textbf{Case~V}, historical samples are used to construct a data-driven uncertainty set, reducing unnecessary conservativeness while still providing protection against adverse realizations.

\subsubsection{Out-of-sample Performance Comparison}
Table~\ref{tabII} reports the out-of-sample performance of \textbf{Cases~III--V} on $\bm\Xi_{\rm out}$ in terms of both cost and credibility. Here, P95 denotes the 95th percentile of the corresponding cost metric over $\bm\Xi_{\rm out}$, meaning that 95\% of the out-of-sample scenarios have costs no greater than the reported P95 value. As shown in Table~\ref{tabII}, \textbf{Case~III} has the highest P95 total cost ($\$12.61\times 10^{6}$), followed by \textbf{Case~IV} ($\$12.33\times 10^{6}$), whereas \textbf{Case~V} has the smallest P95 total cost ($\$11.33\times 10^{6}$). This indicates that the SP-based plan is more vulnerable to adverse realizations, whereas the proposed data-driven RO mitigates tail risk while avoiding the heavy investment implied by the box-robust plan.

In terms of credibility, \textbf{Case~III} achieves the lowest violation probability (0) on $\bm\Xi_{\rm out}$. \textbf{Case~IV} exhibits the highest violation probability (3.0\%), while \textbf{Case~V} reduces it to 0.6\%, providing a clear improvement in reliability over SP.

Overall, \textbf{Case~V} offers a more balanced profit--risk trade-off: it relaxes the conservatism of \textbf{Case~III} by exploiting data-driven uncertainty information, while delivering higher credibility than \textbf{Case~IV}.
\begin{figure}[!t]\vspace{0em}
	\centering
	\includegraphics[width=8.8cm]{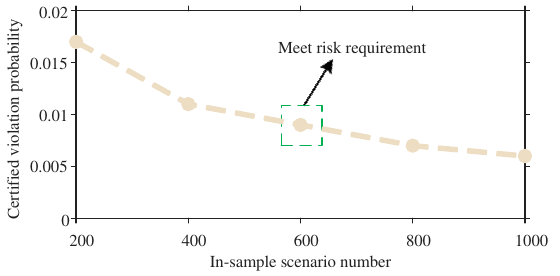}
	\caption{\textit{A posteriori} feasibility guarantee and out-of-sample feasibility.}
	\label{fig9}
\end{figure}
\begin{table}[!t]\vspace{-0em}
	\renewcommand{\arraystretch}{1.08}
	\centering
	\caption{Out-of-sample comparison on the out-of-sample set $\bm\Xi_{\rm out}$}
	\label{tabII}
	\setlength{\tabcolsep}{0.7mm}
	\begin{tabular}{c|c|c|c|c}
		\toprule[0.8pt]
		\multirow{2}{*}{\textbf{Case}} 
		&\textbf{Investment}
		& \multicolumn{1}{c|}{\textbf{Operating cost} ($\$10^{6}$)}
		& \multicolumn{1}{c|}{\textbf{Total cost ($\$10^{6}$)}}& \textbf{Violation}\\
		&\textbf{cost ($\$10^{6}$)}  & {P95}  & {P95}&\textbf{probability}\\
		\hline
		\textbf{III} & 4.60  & 8.01  & 12.61  & 0\% \\
		\textbf{IV}  & 4.10  & 8.23  & 12.33  & 3.0\% \\
		\textbf{V}   & 4.10  & 7.23  & 11.33  & 0.6\% \\
		\bottomrule[0.8pt]
	\end{tabular}
\end{table}

\subsection{Validation of the \textit{A Posteriori} Risk Certificate} 
Fig.~\ref{fig9} shows the certified violation-probability bound $\hat{\varepsilon}$ versus the in-sample scenario size. As the scenario number increases from 200 to 1000, $\hat{\varepsilon}$ drops from 0.017 to about 0.006. The risk bound tightens as more scenarios are included, thereby strengthening the certificate. At the third iteration (scenario number $=600$), the prescribed risk level is already met, i.e., $\hat{\varepsilon} (0.008) \le \bar{\varepsilon} (0.01)$. Moreover, the out-of-sample violation probability (0.6\% in Table~\ref{tabII}) is lower than $\hat{\varepsilon}$. Further enrichment beyond this point continues to reduce the bound and increases the safety margin. These results demonstrate the effectiveness of the proposed \textit{a posteriori} risk certificate and support its use as a practical stopping criterion for scenario enrichment.

\subsection{Profit--Risk Trade-Off: Sensitivity to Risk Tolerance}
\begin{figure}[!t]
	\vspace{-1em}
	\centering
	\includegraphics[width=0.98\columnwidth]{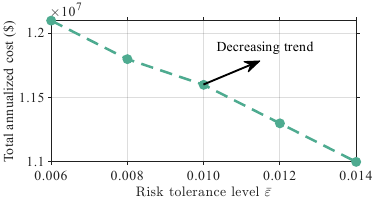}
	\caption{Sensitivity to the risk tolerance $\bar{\varepsilon}$.}
	\label{fig10}	
\end{figure}

The sensitivity analysis based on \textbf{Case~V} is conducted to examine the profit--risk trade-off enabled by the proposed risk-tunable procedure (Fig.~\ref{fig4}). The confidence level is fixed at $\beta=0.05$, and the prescribed risk tolerance level $\bar{\varepsilon}$ is varied. Fig.~\ref{fig10} shows the resulting total annualized cost. When the risk tolerance level $\bar{\varepsilon}$ increases from $0.006$ to $0.014$, the total annualized cost decreases from $\$1.21\times 10^{7}$ to $\$1.10\times 10^{7}$, representing a reduction of about $9.1\%$. These results can assist system operators in formulating operational strategies with an optimal trade-off between profit and risk.

\section{Conclusion}
This paper develops a data-driven two-stage robust model for DWCL planning in CTPNs. An MUE model incorporating battery SoC dynamics and endogenous charging decisions is used to represent EV routing and charging behavior. Joint historical profiles of traffic demand and RES output are used to preserve their spatio-temporal correlations without assuming a parametric distribution. After linearization, the two-stage RO model with mixed-integer recourse is reformulated as an equivalent single-stage SA model and solved using a customized NCCG algorithm. An \textit{a posteriori} feasibility certificate based on the number of support scenarios is developed to quantify out-of-sample risk, and a risk-guided scenario-enrichment procedure is used to meet a prescribed tolerance. Case studies show that endogenous charging decisions reduce annualized cost by 18.7\% relative to the UE model that requires charging whenever a DWCL is traversed. The proposed data-driven RO framework reduces the out-of-sample P95 total cost by 10.2\% and 8.1\% relative to conventional RO and SP, respectively, while limiting the violation probability to 0.6\%. Sensitivity analysis further shows that relaxing the risk tolerance from 0.006 to 0.014 reduces the total annualized cost by 9.1\%, demonstrating the tunable profit--risk trade-off.

Several interesting research directions for future work include: i) extending the proposed framework to multistage planning with nonanticipativity constraints to account for unforeseen events; and ii) developing resilience-oriented planning that considers extreme scenarios.

\appendices
\section{Proof of the Equivalence Between \eqref{I} and \eqref{J}}
\label{appendix_A}
For the two-stage RO model \eqref{I}, since the adversarial maximization in \eqref{I_1} is taken over the finite scenario set
$\bm \Xi = \{\bm \xi^{(s)}\}_{s=1}^{S}$, the worst-case value is attained at one of the $S$ scenarios.

For any fixed $\bm x$, define the scenario-wise value function as
\begin{equation}
	\begin{aligned}
		\phi^{(s)}(\bm x) = \min_{\bm y\in\bm Y,\ \bm z\in\bm Z}\Big\{&\bm d^{\top}\bm y:  \ \bm B\bm x+\bm C\bm \xi^{(s)}+\bm D\bm y\\ 
		& +\bm E\bm z\le \bm h,~ s=1,\ldots,S\Big\}.
	\end{aligned}	\label{Z_1}
\end{equation}
Note that, to avoid infeasibility caused by hard voltage limits, the voltage constraints are softened by allowing bounded violations and adding a sufficiently large penalty term in the objective of \eqref{Z_1}.

Then \eqref{I_1} is equivalent to
\begin{equation}
	\min_{\bm x\in\bm X}\ \bm c^{\top}\bm x+\max_{s=1,\ldots,S}\phi^{(s)}(\bm x).
	\label{Z_2}
\end{equation}

Using the epigraph form of the maximum, we obtain
\begin{equation}
	\max_{s}\phi^{(s)}(\bm x)
	=\min_{\eta\in\mathbb{R}}\ \big\{\eta:\ \eta\ge \phi^{(s)}(\bm x),\ s=1,\ldots,S\big\}\label{Z_3}
\end{equation}

By introducing per-scenario decision copies $(\bm y^{(s)},\bm z^{(s)})$, the above formulation is equivalent to \eqref{J}.

\bibliographystyle{IEEEtran}
\bibliography{ref}
\vfill
\end{document}